\documentclass[aps, pre, showpacs]{revtex4}
\usepackage[colorlinks=true,linkcolor=blue,citecolor=blue]{hyperref}

\usepackage[dvips]{graphicx}
\usepackage{epsfig}
\usepackage{ulem}
\usepackage[T1]{fontenc}
\usepackage[latin9]{inputenc}
\usepackage{esint}
\usepackage[english]{babel}
\usepackage{ulem}
\usepackage{color}
\makeatletter

\begin{document}

\title{Experimental observation of front propagation in LL model
with negative diffractive  and inhomogeneous Kerr cavity}

\author{V. Odent$^{1}$, M. Tlidi$^{1}$,  M.G. Clerc$^{2}$,   and  E. Louvergneaux $^{1}$}
\affiliation{$^{1}$ Laboratoire de Physique des Lasers, Atomes et Mol\'ecules, 
CNRS UMR8523, Universit\'e Lille 1, 59655 Villeneuve d'Ascq Cedex, France}
\affiliation{$^{2}$Facult{\'e} des Sciences, Universit{\'e} libre de Bruxelles (U.L.B.), C.P.
231, Campus Plaine, B-1050 Bruxelles, Belgium}
\affiliation{$^{3}$Departamento de F{\'i}sica, Universidad de Chile, Blanco
Encalada 2008, Santiago, Chile}

\begin{abstract}
A driven resonator with focusing Kerr nonlinearity shows stable localized structures in a 
region far from modulational instability. The stabilization mechanism is based on front interaction 
in bistable regime with an inhomogeneous injected field. The experimental setup consist of a 
focusing Kerr resonator filled with a liquid crystal and operates in negative optical diffraction regime. 
Engineering diffraction is an appealing challenging topic in relation with left-handed materials. 
We solve the visible range of current left-handed materials  to show that localized structures 
in a focusing Kerr Fabry-Perot cavity submitted to negative optical feedback are propagating 
fronts between two stable states. We  evidenced analytically, numerically, and experimentally 
that these fronts stop due to the spatial inhomogeneity induced by the laser Gaussian forcing,
 which changes  spatially the relativity stability between the connected states.
\end{abstract}

\maketitle
\section{Introduction}
\label{sec:1}

Localized structures (LS's) often called cavity solitons in dissipative media have 
been observed in 
various fields of nonlinear science (see last overview on this issue 
\cite{Leblond-Mihalache,Tlidi-PTRA,DCRA2011}).  
Localized structures consist of isolated or randomly distributed peaks surrounded by 
regions in the homogeneous steady state. Currently they attract growing interest in optics 
due to potential applications for all-optical control of light, optical storage, and information 
processing. They appear in an optical resonator containing a third order nonlinear media 
such as liquid crystals, and driven coherently by an injected beam in a positive diffraction 
regime and close to modulational instability \cite{Scroggie,Odent.2011}.  The existence of 
localized structures and localized patterns, due to the occurrence of a modulational instability, 
has been abundantly discussed and is by now fairly well understood \cite{Pomeau}. 
In this case, LS's appears in the subcritical modulational instability regime where there is a coexistence 
between the homogeneous steady state and the spatially periodic pattern. 
Localized structures consist of isolated or randomly distributed spots surrounded 
by regions in the uniform  state. They may consist of peaks or dips embedded in the 
homogeneous background. 

However, localized structures could be formed in modulationally stable regime \cite{Nonturing}. 
In this case, heterogeneous initial conditions usually caused by the inherent fluctuations generate 
spatial domains, which are separated by their respective interfaces often called front solutions or 
interfaces or domain walls \cite{Rev4}. Interfaces between these metastable states appear in the 
form of propagating fronts and give rise to a rich spatiotemporal dynamics \cite{Langer,Collet}.  
From the point of view of dynamical system theory at least in one spatial dimension a front is a 
nonlinear solution that is identified in the comoving frame system as a heteroclinic orbit 
linking two spatially extended states \cite{vanSaarloos,Coullet2002}.
The dynamics of the interface depends on the nature of the states that are connected. 
In the case of a front connecting a stable and an unstable state, it is called as 
Fisher-Kolmogorov-Petrosvky-Piskunov (FKPP) front  \cite{Rev8,Fisher,Kolmogorov,VanSaarloos03}. 
One of the characteristic features of these fronts is that the speed is not unique, nonetheless 
determined by the initial conditions. When the initial condition is bounded, after a transient, 
two counter propagative fronts with the minimum asymptotic speed emerge \cite{Rev8,VanSaarloos03}. 
In case that the nonlinearities are weak, this minimum speed is determined by the linear or marginal-stability 
criterion and fronts are usually referred to as pulled \cite{VanSaarloos03}. 
In the opposite case, the asymptotic speed can only be determined by nonlinear methods 
and fronts are referred to as pushed \cite{VanSaarloos03}.
The above scenario changes completely for a front connecting two stable states. In the case 
of two uniform states, a gradient system tends to develop the most stable state, in order to 
minimize its energy, so that the front always propagates toward the most energetically favored 
state \cite{ClercNagaya04}. It exists only as one point in parameter space 
for which the front is motionless, 
which is usually called the Maxwell point, and is the point for which the two states 
have exactly the same energy for variational systems  \cite{Goldstein}.

In this chapter, we present analytical, numerical and experimental investigation of of front 
interaction in bistable regime with an inhomogeneous injected beam. Far from any modulational 
instability, the inhomogeneous injected beam in the form of  Gaussian, 
can lead to the stabilization of LS's. This completes our previous communication of the stability 
of LS's \cite{Odent-RC}. We consider an  experimental setup which consists of a focusing Kerr 
resonator filled with a liquid crystal and operates in negative optical diffraction regime. 
Engineering diffraction is an appealing challenging topic in relation with left-handed materials. 
We solve the visible range of current left-handed materials 
\cite{Kockaert.2006,Tassin.2006,Tassin.2007,Tlidi.2011} to show that LS's  in a focusing 
Kerr Fabry-Perot cavity submitted to negative optical feedback are propagating fronts 
between two stable states.  Moreover, we have evidenced that these fronts 
stop due to the spatial inhomogeneity induced by the laser Gaussian forcing,
which changes  spatially the relativity stability between the connected states.

This chapter is organized as follows. After an introduction 
we present the experimental setup and experimental observation in Sec. 2. 
The analytical and numerical investigations are presented in Sec 3. 
The special case with a null diffraction is  developed in Sec. 4. We conclude in Sec. 5.

\begin{figure}[tpb]
\centering
\includegraphics[width=16.0 cm]{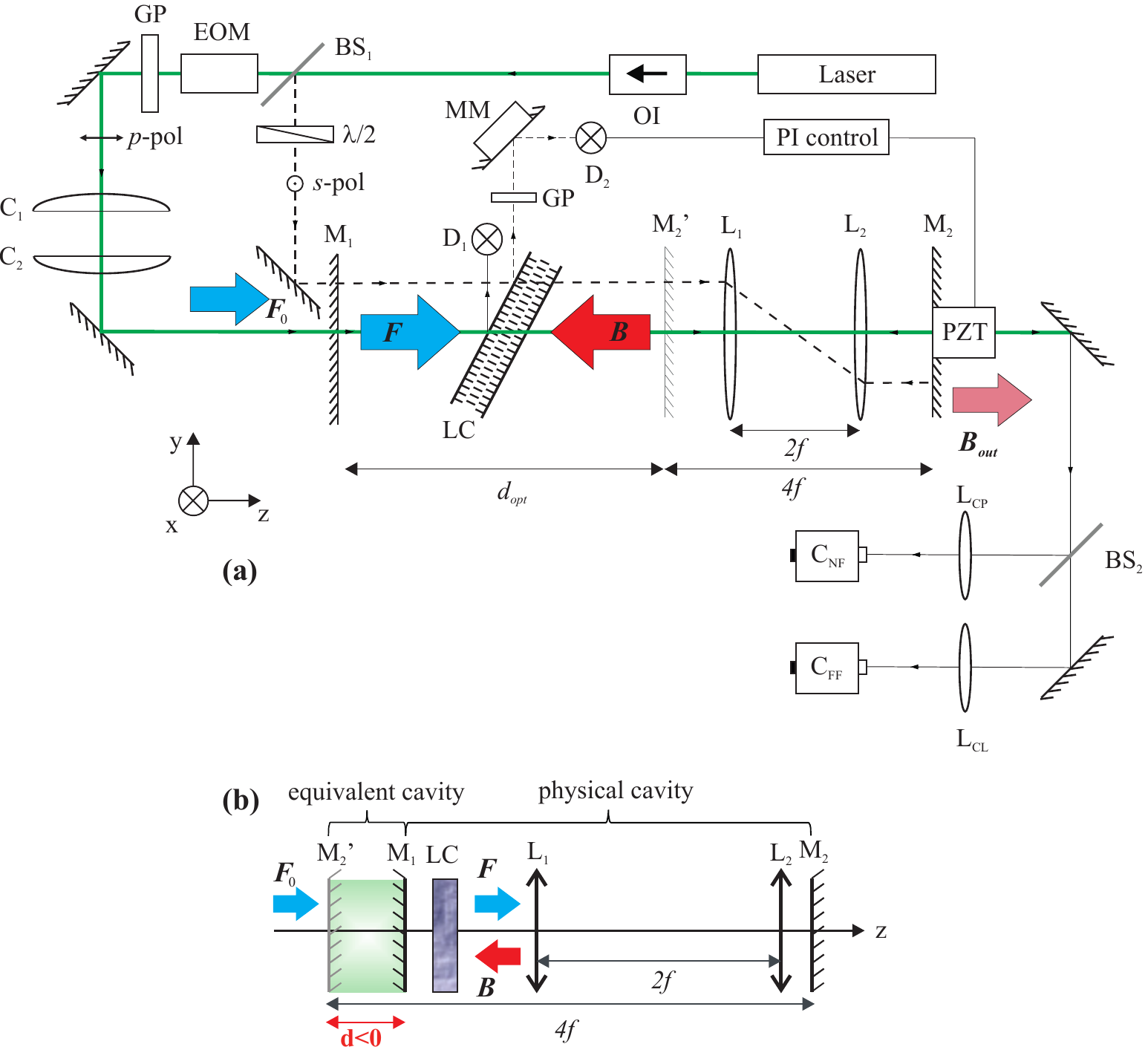} 
\caption{Inhomogeneous Kerr cavity with negative diffraction: (a) Experimental setup. (b) Schematic 
representation of the physical cavity and the equivalent cavity, with negative cavity length. OI optical 
isolator; BS beam splitter; EOM electro-optical modulator; GP Glan polarizer; $C_{1}$ and $C_{1}$ cylindrical lenses;
LC liquid crystal slice; $\textrm{D}{}_{1}$ and $\textrm{D}{}_{2}$ photodetectors; MM motorized mirror;
$\textrm{L}{}_{1}$ and $\textrm{L}{}_{2}$ lenses of focal length
$f$: $p$ and $s$ are the polarized components of the pump (solid
line) and probe (dashed line) beams respectively; $C_{NF}$ and $C_{FF}$ near field and far field cameras; $\textrm{M}{}_{1}$
and $\textrm{M}{}_{2}$ are the real cavity mirrors but the optical
Perot-Fabry cavity is delimited by $\textrm{M}{}_{1}$ and $\textrm{M}'{}_{2}$
mirrors and its length is $d$.}
\label{Fig:setup} 
\end{figure}

\section{Experimental}
\subsection{Set-up}

The experiments have been carried out using a nonlinear Kerr slice
medium inserted in an optical Fabry-Perot resonator. The Kerr focusing
medium is a 50-$\mu m$-thick layer of E7 nematic liquid crystal homeotropically
anchored. Two plane mirrors $\textrm{M}{}_{1}$
and $\textrm{M}{}_{2}$ define the physical cavity but the optical
one is delimited by $\textrm{M}{}_{1}$ and $\textrm{M}'{}_{2}$ (which
is the image of $\textrm{M}{}_{2}$ through the $4f$ lens arrangement)
and its optical length is $d$ {[}Fig.~\ref{Fig:setup}{]}. The intensity
reflexion coefficients of $M_{1}$ and $M_{2}$ mirrors are $R_{1}=81.4$
and $R_{2}=81.8$ respectively so that the cavity finesse is estimated
to 15. The experimental recording of the Airy function (blue curve) on Fig.~\ref{Fig:Phase shift} gives a finesse
of 11.6 indicating the presence
of supplementary losses due to nonlinear medium and lenses transmission
coefficients.

\begin{figure}[tpb]
\begin{centering}
\includegraphics[width=1\columnwidth]{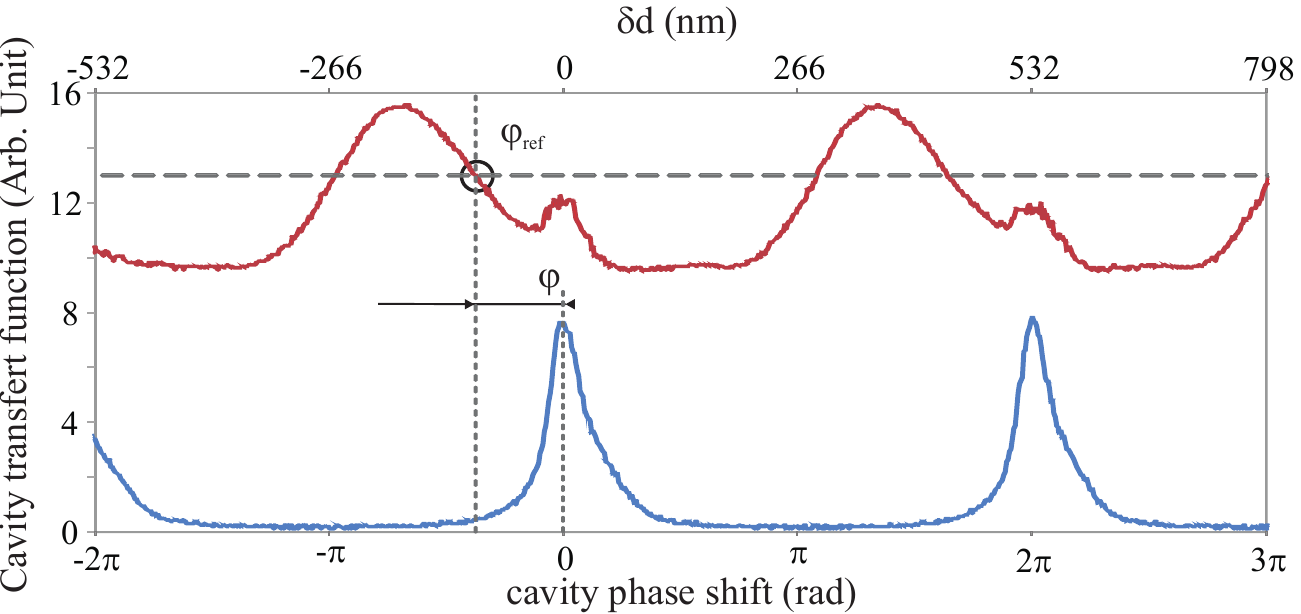} 
\par\end{centering}
\caption{ Transfer function of the optical cavity for : pump beam (blue line) and probe beam (red line).
$\varphi$ linear phase shift of the cavity. $\delta d$ cavity length variation.}
\label{Fig:Phase shift} 
\end{figure}

The optical cavity length $d$ may be tuned from positive to negative values
{[}positive on Fig.~\ref{Fig:setup}(a) and negative on Fig.~\ref{Fig:setup}(b){]}. Thus,  for negative
optical cavity path ($d<0$), a beam propagating along this path experiences
negative diffraction. Together with the positive Kerr index, the
experimental setup is then equivalent to a Kerr cavity that would
have a positive optical distance but negative Kerr index (the $\eta a$
product sign in Ref. \cite{LL} that defines the type of transverse
instabilities remains the same). However the physical mechanisms of negative refraction 
and negative diffraction are different. Thus, this intra-cavity geometrical
lens arrangement allows for achieving an equivalent left-handed Kerr
material in the visible range. It also allows to continuously tune
the diffraction from positive to negative.

\begin{figure}[tpb]
\begin{centering}
\includegraphics[width=1\columnwidth]{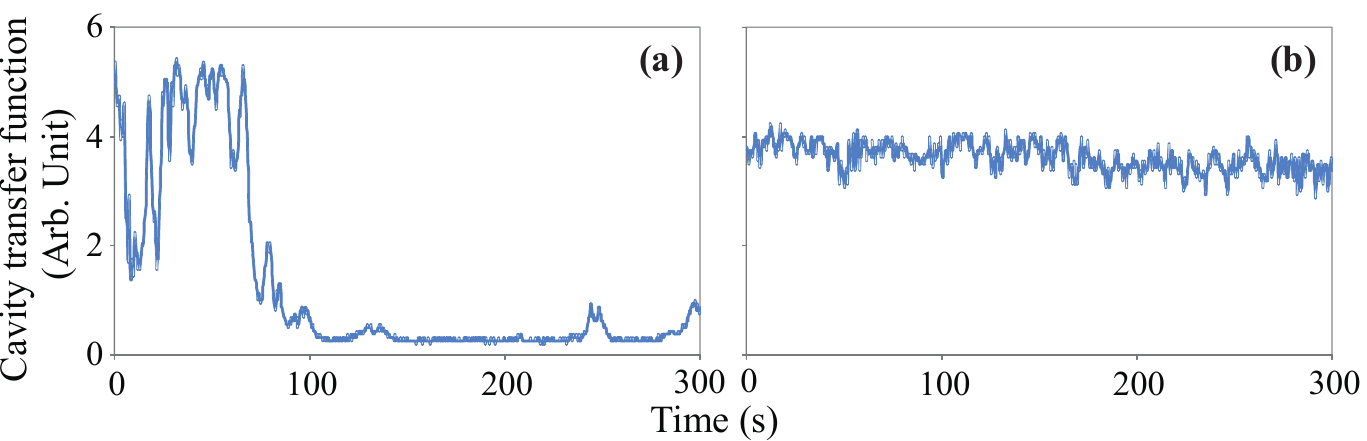} 
\par\end{centering}
\caption{Temporal evolution of cavity transfer function correlated to cavity phase shift: (a) without stabilization; 
(b) with stabilization. $d=5\:mm$, $I=37\:W/cm^{2}$.}
\label{Fig:Phase shift evolution} 
\end{figure}

The laser source used is a single mode frequency
doubled $Nd^{3+}$:YVO4 laser $(\lambda_{0}=532\, nm$) at $5W$.
We have split the laser beam into two parts, a pump beam containing $95\%$ of the initial 
power and a probe beam with $5\%$ of the initial power. The pump beam power is controlled by an electro-optical modulator 
associated with a Glan polarizer. Then it is shaped
by means of two cylindrical telescopes. The resulting beam size ($\sim200\,\mu m\times2800\,\mu m$)
gives a "cigar'' transverse shape such that only one spot can develop
in one of the two directions. This beam is injected inside the cavity may then be considered as
one-dimensional. 
The probe beam polarization is rotated to $90\text{\textdegree}$ to prevent interference phenomenon inside the cavity between the two beams (see Fig.~(\ref{Fig:setup}))
The pump beam propagating in the forward direction is
monitored at the output of $\textrm{M}{}_{2}$ mirror. We record the near field on  $C_{NF}$ and 
the far field on $C_{FF}$.
The cavity detuning is an important parameter, because it determines the solution type (mono/bistable) of the cavity and also the energy quantity inside the cavity, as shown on the cavity transfer 
function on Fig.~\ref{Fig:Phase shift}. For these reasons, we perform an active stabilization of the cavity phase shift, based on Coen's work and coworkers \cite{Coen.1999}. We stabilize the cavity length around a reference phase on the beam probe ($\varphi_{ref}$ on Fig.~\ref{Fig:Phase shift}), then we measure the phase shift between the probe and the pump beams to have the real cavity detuning $\varphi$, as presented on Fig.~(\ref{Fig:Phase shift}). Our setup allow to maintain the detuning long enough to realize experiments, it means to have a fix value $\pm\pi/19$ during many minutes. Without the stabilization, the cavity detuning can evolve quickly and not permit us to do experiments. The typical detuning evolution without and with an active stabilization are presented on Fig.~(\ref{Fig:Phase shift evolution}).

\subsection{Experimental observations}

\begin{figure}[tpb]
\begin{centering}
\includegraphics[width=0.8\columnwidth]{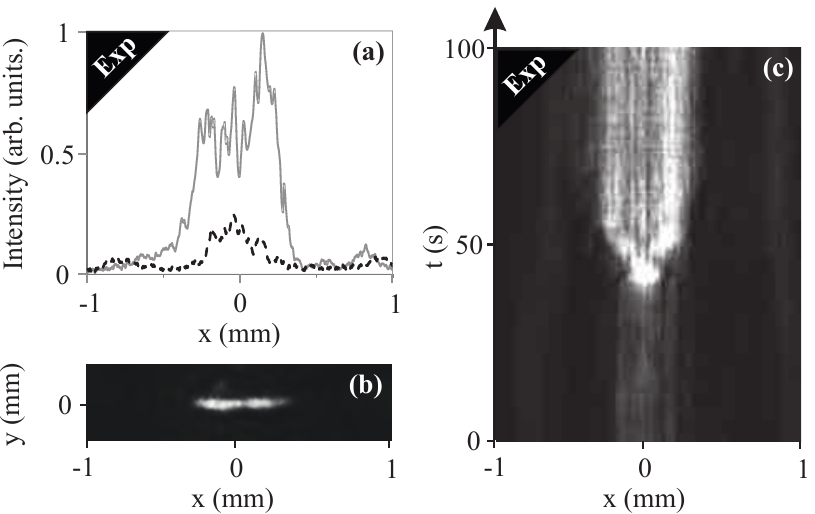} 
\par\end{centering}
\caption{Experimental front propagation in negative diffractive inhomogeneous Kerr cavity.
(a) Transverse cross-section of the initial (final) average localized
structure  in dashed black line (in
continue gray line). (c) Spatiotemporal response to a step function
of the input intensity from the lower to the upper branch of the bistable
cycle. (b) cross section of the localized state. Light region account for high intensity of the light. $I_{0}=433\: W.cm^{-2}$,
$d=-5\: mm$, $\varphi=-0.6$~rad, $w_{x}=1400\:\mu m$,
$w_{y}=100\:\mu m$, $R_{1}=81.8\:\%$; $R_{2}=81.4\:\%$.}
\label{Fig:Experimental results} 
\end{figure}

As the input power is suddenly increased
to the upper bistability response branch ($t=0$~\emph{s}), the central part of transmitted
intensity profile suddenly jumps after some latency time {[}$t \approx 37$~\emph{s}
on Fig.~\ref{Fig:Experimental results}(c){]} to a higher value and invades
the surroundings towards the external regions where the field is less
intense before stopping its propagation. Finally, the fronts locks to give a localized
light state {[}Fig.~\ref{Fig:Experimental results}(b){]}.  Changing the waist $w$ of the Gaussian forcing
or its intensity within the bistability region allows to tune the
distance between the bounded fronts and so the localized state extension.

A transverse cut of the transmitted intensity profile is depicted
on Fig.~\ref{Fig:Experimental results}(a) obtained in the initial and the final observation period.
This figure emphasizes the coexistence of different states in the same region of 
parameters. In addition, a state emerges from the other because of the inherent fluctuations
of the system. Figure~\ref{Fig:Experimental results}(c) shows this phenomenon starting 
at $t \approx 37$~\emph{s}.
From this instant, the system exhibits two counter-propagating fronts, 
which will become asymptotically motionless [see Fig.~\ref{Fig:Experimental results}(c)]. 
The motionless front is observed at the location $x_{0}(expt)\cong\pm0.17\: w$.
At this location, the input intensity is only $3$ \% lower than at the
center of the input beam. 
Thus, we have observed experimentally coexistence between two inhomogeneous states, 
the noise induces a pair of fronts among these states, which initially are counter-propagate
and asymptotically they stop.

\section{Analytical results}
\label{subsec:2}
\subsection{Model equation}
The dynamics of the single  longitudinal mode of the bistable system which consist of a Fabry-Perot cavity  filled with liquid crystal Kerr like medium and driven by a coherent plane-wave 
steady  can be described by the simple partial differential equation 
(the LL model \cite{LL}) in which we incorporate an an inhomogeneous injected field. This mean field approach model is valid under the 
following approximations: ~the cavity possess a high Fresnel number \textit{i.e.}, large-aspect-ratio system and 
we assume that the cavity is much shorter than the diffraction and the nonlinearity 
spatial scales; and ~for the sake of simplicity, we assume a single longitudinal mode 
operation. Under these assumptions the space-time evolution of the
intracavity  field is described by the following partial differential equation 
 \begin{equation}
\frac{\partial E}{\partial t}=E_{in}(x)-(1+i\Delta)E +i|E|^2E-i\left|\alpha\right|\frac{\partial^{2}E}{\partial r^{2}}+\sqrt{\varepsilon}\xi\left(x,t\right)
\label{EqLL}
\end{equation} 
which includes the effect of diffraction, which is proportional to $\alpha$; $E$ is the normalized slowly-varying 
envelope of the electric field, $\Delta$ is the detuning parameter, $E_{in}$ is the 
input field assumed to be real, positive and spatially inhomogeneous. The negative diffraction coefficient is $|\alpha|$. Note that  the above model has been derived for a cavity filled with  left-handed material operating in negative diffraction regime \cite{Tassin.2006}. $\varepsilon$ scales the noise amplitude and $\xi(x, t)$ are Gaussian stochastic processes of zero mean and delta correlation introduced to model thermal noise \cite{Agez2005}.

\subsection{Derivation of F-KPP equation}

The following development is realized for the deterministic case of Eq.~\ref{EqLL} ($\varepsilon=0$). The homogeneous steady states of Eq.~(\ref{EqLL}) are solutions of $E_i=[1+i(\Delta -|E_{s}|^{2})]E _{s}$. The response curve involving the intracavity intensity $\vert{E_{s}}\vert^{2}$ as a function of the input intensity $\vert{E_{i}}\vert^{2}$ is monostable for $\Delta<\sqrt{3}$  and exhibits a bistable behavior when the detuning $\Delta>\sqrt{3}$. For  $\Delta=\sqrt{3}$, we obtain the critical point associated with bistability where the output versus input characteristics have an infinite slope. At the critical point, the coordinate of the intracavity are $E_c=u_c+iv_c$ with $u_c=3^{1/4}/\sqrt{2}$ and $v_c=-1/3^{1/4}\sqrt{2}$, and the injected field amplitude is $E_{in}^c=2\sqrt{2}/3^{3/4}$.  

The analytical investigation of fronts  dynamics  connecting two-homogeneous steady states in 
the framework of the Lugiato-Lefever model Eq.~(\ref{EqLL}) is far from the 
scope of the present chapter. In this section 
we perform a derivation  of  a simple bistable model  with 
inhomogeneous injection to study analytically the dynamics of the front connecting the two-homogeneous steady states. To do that,  
we introduce a small parameter that measures the distance from the critical point $\zeta \ll 1$ 
and we express the cavity detuning in the form 
\begin{equation}
\Delta=\Delta_c(1+\zeta ^2\sigma)
\end{equation}
 where $\sigma$ is a quantity of order one. Then, we decompose the envelope of the electric field into its  real and imaginary parts: $E=x_1+ix_2$
and we introduce a new space and time scales as $(x,t)$  $\equiv$ $\left(\zeta^2t/\sigma,3^{\frac{1}{4}}\zeta x/(\sqrt{\sigma})\right)$. The injected field can be expanded as
\begin{equation}
E_{in}=E_{in}^c\left(1+\frac{3\zeta ^2I}{4}+\zeta ^3\alpha+...\right)
\end{equation}

Let ($u, v)= (x_1, x_2)- (u_c,v_c)$ be the deviations of the real, the imaginary parts of 
intracavity field with respect to the values of these quantities at the critical point. 
Inserting these expansion  into the LL model and using the above scalings, we obtain

\begin{eqnarray}
\frac{\partial u}{\partial t}&=&\zeta\left(\frac{u^2}{2}+uv+\frac{v^2}{2}\right)+\zeta^2\left(\frac{4\alpha}{3}-Iv+\frac{u^2v}{2}+\frac{v^3}{6}\right)-\frac{1}{\sqrt{3}} \frac{\partial^{2}v}{\partial x^{2}},\\
\frac{\partial v}{\partial t}&=&-2(u+v)+\zeta\left(3I-\frac{9u^2}{2}-uv-\frac{v^2}{2}\right)+\frac{\zeta^2}{2}(6Iu-3u^2-uv^2)+\sqrt{3}\frac{\partial^{2}u}{\partial x^{2}}.
\end{eqnarray}

Our aim is to seek solutions of Eqs.~(4,5) in the neighborhood of the critical 
point associated with the optical  bistability. To this end, we expand the cavity field and the injected field as 
\begin{equation}
(u, v)=\zeta [(u_0,v_0)+\zeta (u_1,v_1)+\zeta^2 (u_2,v_2)+\zeta^3 (u_3,v_3)+...]
\end{equation}
Inserting these expansions and taking into account of Eqs. (2,3)  into the LL 
model and using the above scalings, we then obtain a hierarchy of linear problems for the 
unknown functions. At the first order in $\zeta$, we find $u_0=-v_0$. At the second, we 
have $v_1=-u_1+3I/2-u_0^2$. Finally, at the third order, we get a simple bistable model
\begin{equation}
\frac{\partial u}{\partial t}=f(u)+\frac{\partial^{2}u}{\partial x^{2}}
\label{eq:generic_model}
\end{equation}
where $u(x,t)=\sqrt{3/(2\sigma)}u_0 $ is a scalar field that accounts for the real part of the envelope $E$,
\begin{equation}
f(u)=\eta+u-u^{3}
\label{eq:imperfect_bifurcation}
\end{equation}
and $\eta=4y_2\sigma/3$ controls the relative stability between the equilibria.
Note that $y_2$ is proportional to the pumping $E_{in}$. Hence, 
if the pumping is inhomogeneous then the parameter $\eta$ is also inhomogeneous.
For a Gaussian pumping, we consider  
\begin{equation}
\eta(x)\equiv \tilde{\eta}+\eta_0e^{-(x/w)^2},  
\end{equation}
where $\eta_{0}$ accounts for the strength of the spatial pumping beam
and $w$ is the width of the Gaussian. 
For $\eta(x)=0$ both states are symmetric  corresponding to the Maxwell point, where a 
front between these states is motionless.

To perform analytical developments, we use further approximation by taking into account only the  
first order development close to the center of the optical pumping
where the stress is maximum, i.e., 
\begin{equation}
\eta(x) \approx-\tilde\eta+\eta_{0}\left(1-(x/w)^{2}\right)
\end{equation}
close to the Maxwell point, one can consider the following ansatz for the  front solution 
$u(x,t)=\tanh\left[(x-x_0(t))/\sqrt{2} \right]+H$, 
where $x_{0}$ is the front position and  $H$ accounts for small corrections.  
We  substitute the above ansatz for $u(x,t)$ in Eq.~(\ref{eq:generic_model}), 
linearizing in $H$ and imposing the  solvability condition, we obtain the 
kinetic equation for the evolution of the front position

\begin{equation}
\dot{x}_{0}=\frac{-3\sqrt{2}}{2}\left[-\tilde\eta+\eta_{0}\left(1-\left(\frac{x_{0}}{w}\right)^{2}-\left(\frac{\pi}{\sqrt{6}w}\right)^{2}\right)\right].
\label{speed}
\end{equation}
this equation takes into account of corrections imputable to inhomogeneities of the injected signal. The  stationary solutions of this equation reads
\begin{equation}
x_0=\pm w \sqrt{1-\frac{\tilde\eta}{\eta_{0}}-\frac{\pi^2}{6w^2}},
\end{equation}
the term $\pi^2/(6w)$ is negligible for $w$ large. In this case, we have 

\begin{equation}
x_0=\pm w \sqrt{1-\frac{\tilde\eta}{\eta_{0}}}.
\end{equation}

\begin{figure}[tpb]
\begin{centering}
\includegraphics[width=1\columnwidth]{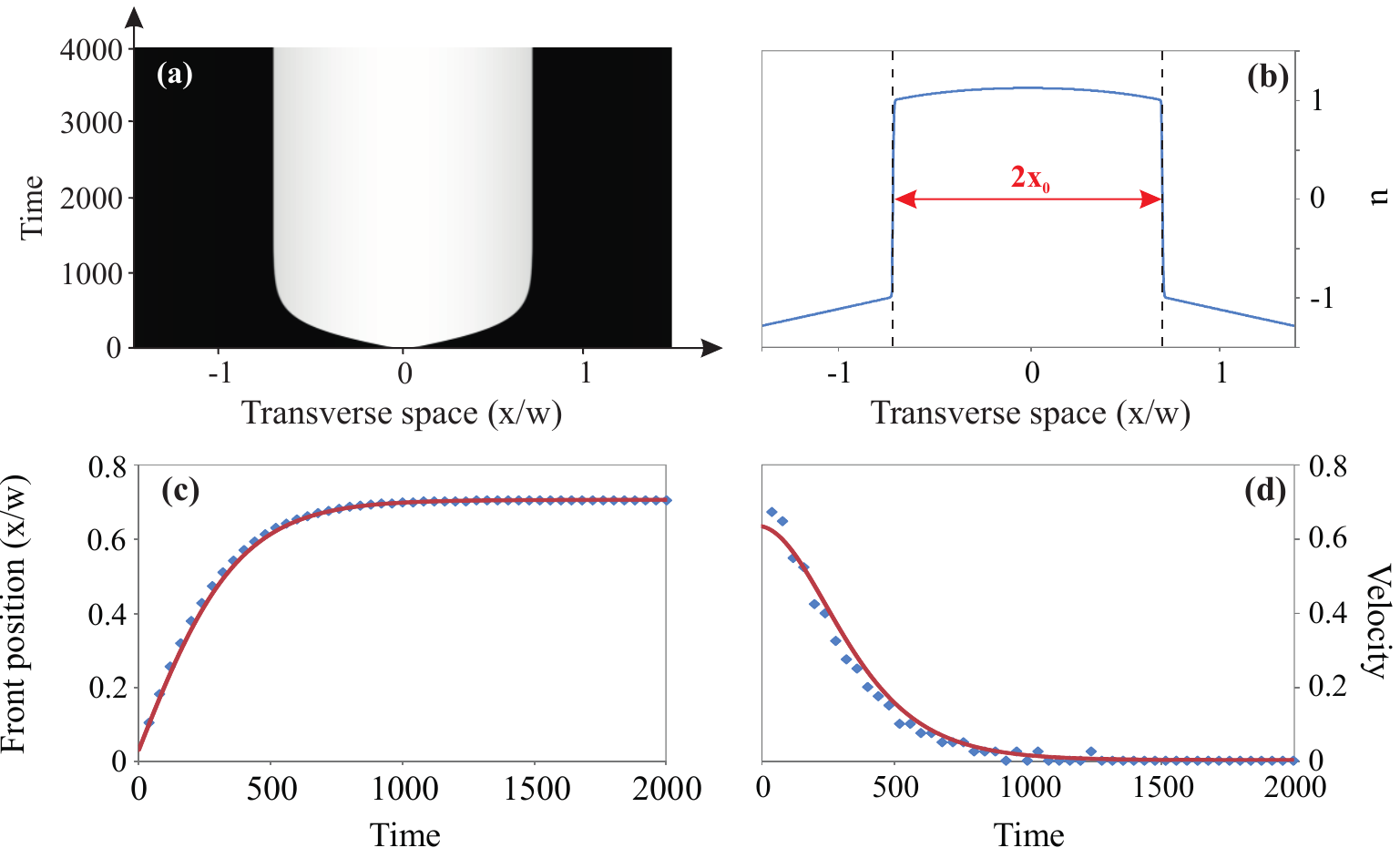} 
\par\end{centering}
\caption{Numerical study of front pinning with a parabolic forcing.
(a) Spatiotemporal diagram; (b) Transverse profile of the front at $t=4000$;
(c) Relative front position $x_{0}(t)$ (for $x>0$); (d) front velocity.
Blue diamonds are the numerical points and the red curves are the
analytical predictions.$\eta_{0}=0.6$, $\tilde\eta=0.3$, $w=350$.}
\label{Fig:SpatioTemporal-Simu} 
\end{figure}

The ordinary differential equation (\ref{speed}) admits an exact solution 
that leads to the following trajectory of the front 

\begin{equation}
x_{0}(t)=\pm a\:\mathrm{tanh}\left(b(t-t_{0})\right),
\label{eq:x0(t)}
\end{equation}
with \emph{a}, \emph{b} and $t_{0}$ are coefficients depending on
$\eta_{0}$, $\tilde\eta$ and $w$. The equilibrium position of the front $x_{0\infty}$
can be inferred from expression Eq.~(\ref{eq:x0(t)}) for $t\rightarrow\infty$
as $x_{0\infty}=\pm w\:\sqrt{1-\tilde\eta/\eta_{0}}$. This equilibrium positions
could be obtained directly from Eq.~(\ref{eq:generic_model}), assuming $\eta(x_{0\infty})=0$
which is the condition for motionless front (Maxwell
point). Extending this last
property to the initial Gaussian forcing,
we get 
\begin{equation}
x_{0\infty}=\pm w\:\sqrt{ln\left(\frac{\eta_{0}}{\tilde\eta}\right)}.
\label{eq:x0inf}
\end{equation}
At leading order, Eq.~(\ref{eq:x0inf}) recover again the
previous expression of $x_{0\infty}$ for the parabolic approximation
of the Gaussian profile.

\subsection{Numerical results}

\begin{figure}[tpb]
\begin{centering}
\includegraphics[width=0.8\columnwidth]{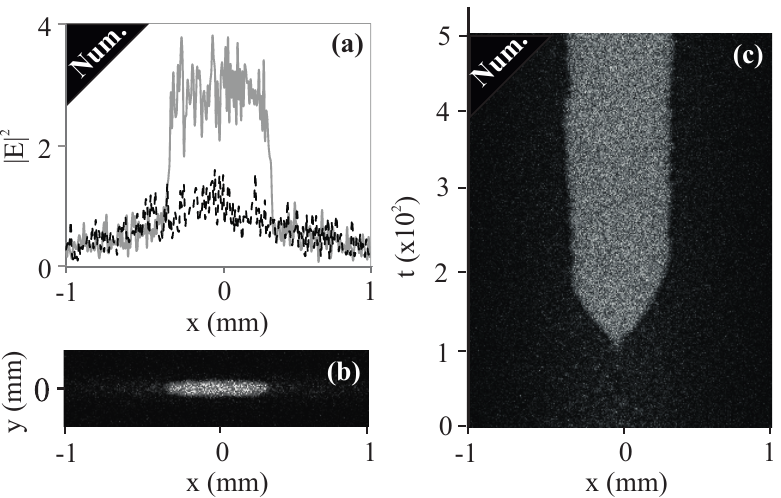} 
\par\end{centering}
\caption{Front propagation in negative diffractive inhomogeneous Kerr cavity.
(a,b) Transverse cross-section of the initial (final) average localized
structure  in dashed black line (in
continue gray line). (c,d) Spatiotemporal response to a step function
of the input intensity from the lower to the upper branch of the bistable
cycle. (e,f) Experimental and numerical cross section of the localized state. 
Light region account for high intensity of the light.(a,c,e) Experiments $I_{0}=433\: W.cm^{-2}$,
$d=-5\: mm$, $\varphi=-0.6$~rad, $w_{x}=1400\:\mu m$,
$w_{y}=100\:\mu m$, $R_{1}=81.8\:\%$; $R_{2}=81.4\:\%$.
(b,d) Numerical simulation of LL model with $E_{0}=1.9$, $\Delta=3.0$, 
$\alpha=0.001$, $w_{x}=1400\:\mu m$,
$w_{y}=100\:\ \mu m$, $\varepsilon=0.4$.}

\label{Fig: Numerical Simulations LL} 
\end{figure}

We  conduct numerical simulations of the imperfect pitchfork bifurcation, 
Eq.~(\ref{eq:generic_model}-\ref{eq:imperfect_bifurcation}), with a parabolic spatial 
injection to compare with the analytical predictions. Initially, we observe the 
front propagation, as the case with a plane wave. Then, the front speed decreases to become zero,
around $t = 1000$, as presented on [Fig.~\ref{Fig:SpatioTemporal-Simu}(a,c,d)]. At $t = 4000$, 
the structure is pinned by the pump parabolic profile and 
presents a bistable structure with a parabolic dependance on the two states [Fig.~\ref{Fig:SpatioTemporal-Simu}(b)]. 
We plot on the same graph [Fig.~\ref{Fig:SpatioTemporal-Simu}(c,d)], 
the analytical predictions of the core front position and its speed 
with the numerical simulations. We have an excellent agreement between both. 
For the Gaussian case, which is not presented in this document, the hyperbolic tangent allows always to reproduce the front trajectory. 
We need only to adjust the parameters $a$ and $b$ from Eq.~(\ref{eq:x0(t)}) to have a good agreement between 
the predictions and the numerical simulations.

For the LL model, Eq.~(1), we focus our numerical investigations with a bi-dimensional Gaussian spatial injection  and by taking into account 
the experimental parameters are shown in Fig.~\ref{Fig: Numerical Simulations LL}.

\begin{figure}[tpb]
\begin{centering}
\includegraphics{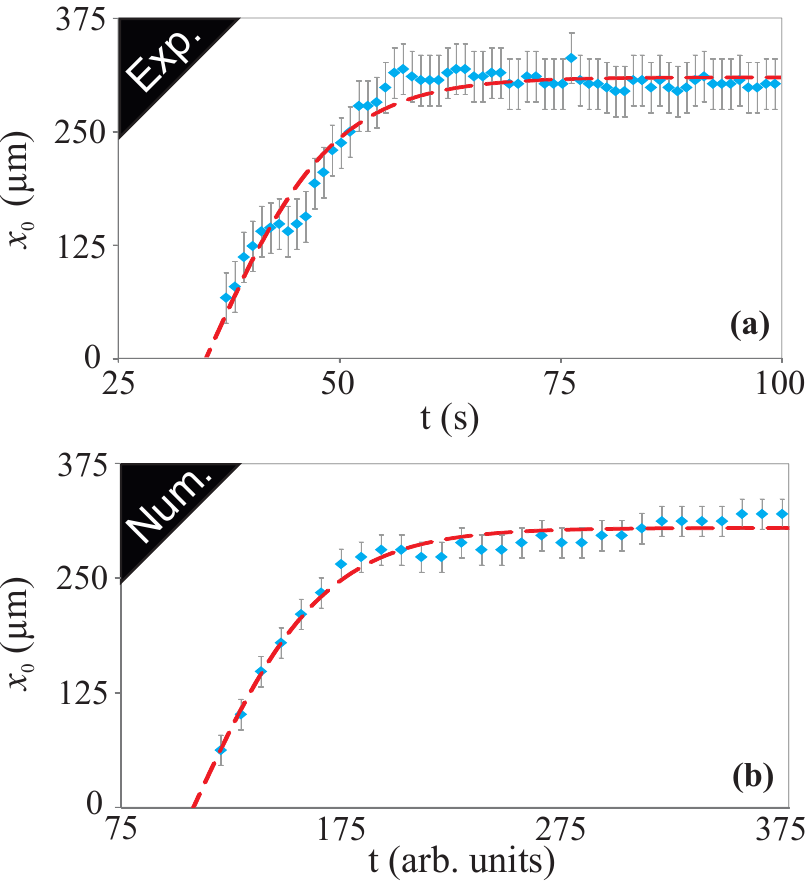}
\par\end{centering}
\caption{Temporal evolution of front  position $x_{0}(t)$ corresponding to
spatiotemporal diagrams of Fig.~(\ref{Fig:Experimental results}-\ref{Fig: Numerical Simulations LL}). (a) Experiment
and (b) numerical simulations of the LL model, Eq.~(\ref{EqLL}), used the 
same parameters considered in Fig.~(\ref{Fig:Experimental results}-\ref{Fig: Numerical Simulations LL}). 
Blue diamonds : location values extracted
from the smoothed spatiotemporal diagrams. Dashed red curves: best fit
using expression~\ref{eq:x0(t)}. Experimental fit parameters : $a=308 \,\mu m$, $b=0.069$, $t_0=35.2\, s$, numerical fit parameters : $a=306\, \mu m$, $b=0.017$, $t_0=107.7\, s$.}
\label{Fig:MI-soliton} 
\end{figure}

We perform numerical simulations 
with a asymmetric Gaussian
forcing  with cigar shape ($w_x \gg w_y$)
\begin{equation}
E_{in}(x,y)=E_0 e^{-\left(\left(\frac{x}{w_x}\right)^2+\left(\frac{y}{w_y}\right)^2\right)}.
\end{equation}
Furthermore, we  take account the inherent fluctuations of the system by the stochastic Lugiato-Lefever model described by Eq.~\ref{EqLL}, $\varepsilon\neq0$. 
We use a stochastic Runge-Kutta solver of the order of 2 with additive noise \cite{Honeycutt.1992}. In this latter case, the temporal step ($\Delta t$) is equal to 0.01.
Numerical simulations with these ingredients show a quite good  agreement
with the experimental observations (see Fig.~\ref{Fig:Experimental results} and Fig.~\ref{Fig: Numerical Simulations LL}).
Hence, 
the analytical expression (\ref{eq:x0(t)}) can be
used to figure out and to characterize the experimental front dynamics. Figure (\ref{Fig:MI-soliton})
depicts the experimental and numerical temporal evolutions of the  front position. It clearly evidences that the expression of Eq.~(\ref{eq:x0(t)})
reproduces well the front dynamics. Therefore, the effect of a spatial
forcing on front propagation is to induce the front moves and stops on an 
asymptotic position, satisfying a hyperbolic tangent trajectory.

\section{ Zero diffraction cavity}

We have explored an interesting limit case, where the diffraction is zero. 
This limit case delimits the positive diffraction case where exists soliton solution {\cite{LL,Odent.2011} and the negative diffraction, 
with front propagation \cite{Odent-RC}. 
It is natural to wonder which kind of solution exists in this limit. 
A linear stability study inform us that the two branch of the hysteresis cycle are linearly stable, contrary to the diffraction cases \cite{LL,Kockaert.2006}. 
Consequently, we expect to observe front solution, connecting the two branch of the bistable cycle. 

The numerical simulations from Eq.~(\ref{EqLL}) have been performed for a  zero  cavity 
length, i.e.  zero  diffraction, reveal the presence of the two states, notwithstanding 
without front propagation [cf. Fig.~\ref{Fig: Zero Diffraction}(a)]. However the experiments 
realized for a  zero  diffraction cavity show two fronts, connecting the dark state with the 
bright state, with slow, asymmetric and irregular speeds [see Fig.~\ref{Fig: Zero Diffraction}(c)]. We conclude that the 
Lugiato-Lefever model is not enough to model the experimental observations for zero diffraction. 
Close to $\alpha=0$, the diffusion of the liquid crystal molecules cannot be neglected. Consequently, we complete 
the model by a second equation governing the spatial nonlinear refractive 
index evolution. the intracavity field $E$ and the nonlinear refractive $n$ are given by

\begin{equation}
\frac{\partial E}{\partial t}=E_{in}(x)-(1+i\Delta)E +inE-i\left|\alpha\right|\frac{\partial^{2}E}{\partial r^{2}}+\sqrt{\varepsilon}\xi\left(x,t\right)
\label{EqLL+N},
\end{equation} 

\begin{equation}
n - \sigma\frac{\partial^{2}n}{\partial r^{2}} = \left|E\right|^{2},
\label{Eq N}
\end{equation} 
where $\sigma$ is the diffusion coefficient. The stochastic numerical simulations realized 
with this model show a good agreement with the experimental observations, as presented on Fig.~\ref{Fig: Zero Diffraction}.

\begin{figure}[tpb]
\begin{centering}
\includegraphics{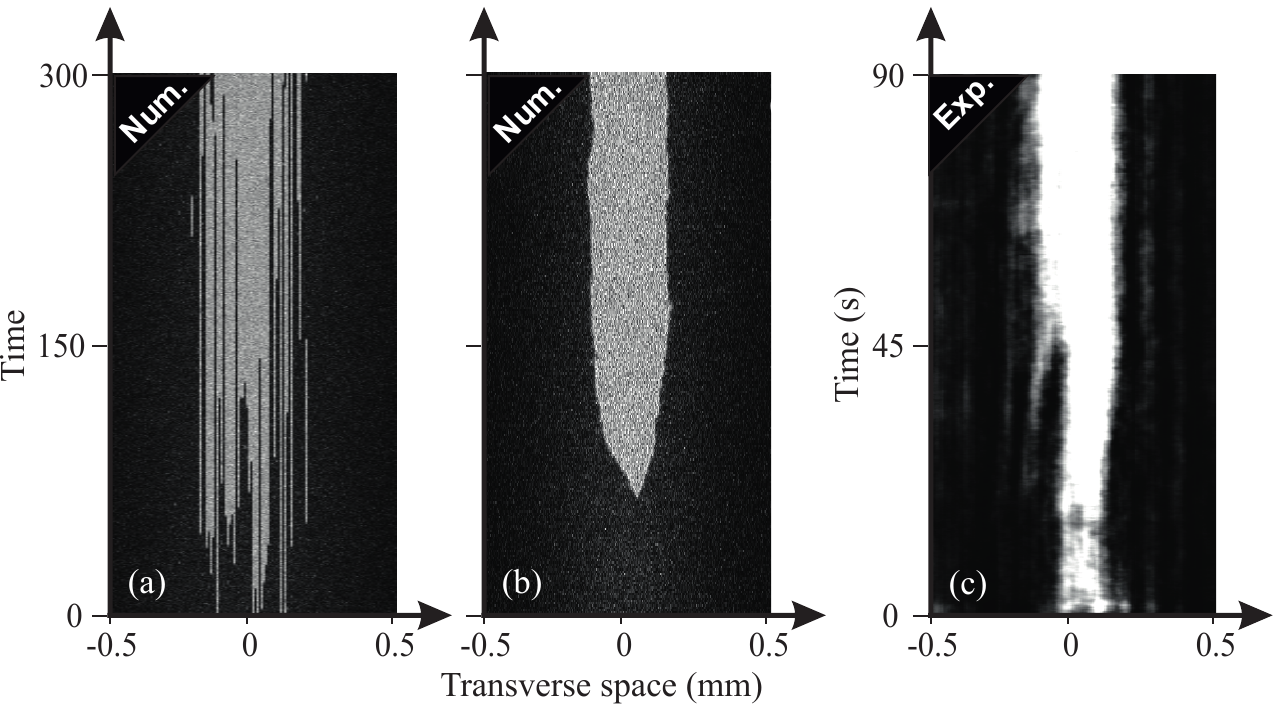}
\par\end{centering}
\caption{Spatiotemporal diagrams: (a) 1D LL model, (b) 1D LL model + liquid crystal diffusion, (c) Experiments.
Experiments $I_{0}=455\: W.cm^{-2}$,
$d=0\: mm$, $\varphi=-0.6$~rad, $w_{x}=1400\:\mu m$,
$w_{y}=100\:\mu m$, $R_{1}=81.8\:\%$; $R_{2}=81.4\:\%$.
Numerical simulation of LL model with $E_{0}=1.95$, $\Delta=3.0$, 
$\alpha=0$, $w_{x}=1400\:\mu m$,, $\varepsilon=0.4$.
}
\label{Fig: Zero Diffraction} 
\end{figure}

\section{conclusion}

We have investigated the formation of localized structures in the Lugiato-Lefever equation in a negative diffraction regime with an inhomogeneous injected beam. We first show how to generate an equivalent left-handed Kerr material in the visible range.
Experimentally, we show that the nonlinear dynamical states appearing
in a focusing Kerr Fabry-Perot cavity submitted to negative optical
feedback are propagating fronts in inhomogeneous medium.

We have performed a reduction of the Lugiato-Lefever equation 
to a simple bistable model  with inhomogeneous injected beam. 
From this simple model we have derived a simple expression for the speed front.  We have used the inhomogeneous spatial pumping in the form of Gaussian beam.
The front moves and stops  on a asymptotic position. The experimental trajectory of the front position under that forcing follows an hyperbolic tangent law that fully agrees with the prediction
from a generic bistable imperfect pitchfork bifurcation model.

Our analysis should be applicable to all fiber resonator \cite{Coen.1999}  with an inhomogeneous injected beam. In this case the coupling is provided by dispersion. When dispersion and diffraction have a comparable influence, three dimensional localized structures can be generated \cite{k48,k47,k48-1,k48-2,k48-3,k49,k48-4}. These structures consist of regular 3D lattices or localized  bright light bullet traveling at the group velocity of light in the material. We plan to extend our analysis to  three dimensional cavities with an inhomogeneous injected beam. In this case,  localized light bullet in the negative diffraction and in anomalous dispersion may be stable.

The authors  acknowledge
financial support by the ANR International program, project no. ANR-
2010-INTB-402-02 (ANRCONICYT39), `COLORS'. 
M. G. thanks for the financial support of FONDECYT project 1150507. This research was also
supported in part by the Centre National de la Recherche Scientifique
(CNRS), the 'Fonds Europ\'een de D\'eveloppement Economique de R\'egions'
and by the Interuniversity Attraction 463 Poles program of the Belgian
Science Policy Office, under 464 Grant No. IAP P7-35 Photonics@be.

\end{document}